\documentclass[twocolumn, aps, pra, superscriptaddress, longbibliography, reprint]{revtex4-1}
\usepackage{graphicx}
\usepackage{amsfonts}
\usepackage{amssymb}
\usepackage{amsthm}
\usepackage{array}
\usepackage{amsmath}
\usepackage{verbatim}
\usepackage{hyperref}
\usepackage{color}
\usepackage{bbold}
\usepackage{epstopdf}
\usepackage{mathtools}
\usepackage[normalem]{ulem}
\usepackage[utf8]{inputenc}

\newcommand{\ket}[1]{\left\vert#1\right\rangle}
\newcommand{\bra}[1]{\left\langle#1\right\vert}

\def\bra#1{ \langle {#1} |}
\def\ket#1{ | {#1} \rangle}

\begin{document}

\title{
Field-gradient measurement using a Stern-Gerlach atomic interferometer with butterfly Geometry
}
\author{Changhun Oh}
\affiliation{Pritzker School of Molecular Engineering, The University of Chicago, Chicago, Illinois 60637, USA}
\author{Hyukjoon Kwon}
\affiliation{QOLS, Blackett Laboratory, Imperial College London, London SW7 2AZ, United Kingdom}
\author{Liang Jiang}
\affiliation{Pritzker School of Molecular Engineering, The University of Chicago, Chicago, Illinois 60637, USA}
\author{M. S. Kim}
\affiliation{QOLS, Blackett Laboratory, Imperial College London, London SW7 2AZ, United Kingdom}
\affiliation{Korea Institute for Advanced Study, Seoul 02455, Korea.}
\begin{abstract}
Atomic interferometers have been studied as a promising device for precise sensing of external fields. Among various configurations, a particular configuration with a butterfly-shaped geometry has been designed to sensitively probe field gradients. We introduce a Stern-Gerlach (SG) butterfly interferometer by incorporating magnetic field in the conventional butterfly-shaped configuration. Atomic trajectories of the interferometer can be flexibly adjusted by controlling magnetic fields to increase the sensitivity of the interferometer, while the conventional butterfly interferometer using Raman transitions can be understood as a special case. We also show that the SG interferometer can keep high contrast against a misalignment in position and momentum caused by the field gradient.
\end{abstract}
\maketitle

\section{Introduction}
Atomic interferometry is a state-of-the-art technique that manipulates a coherent superposition of atomic states for precise field sensing tasks \cite{cronin2009optics}. Among various types of atomic interferometers, an atomic fountain configuration using stimulated Raman transitions \cite{kasevich1991atomic, kasevich1992measurement} has attracted much attention due to the high sensitivity and has been employed for a precise measurement of the acceleration of gravity \cite{peters1999measurement, fray2004atomic, muller2008atom-2, hu2013demonstration, schlippert2014quantum}, the Newtonian gravitational constant $G$ \cite{fixler2007atom, lamporesi2008determination, rosi2014precision, prevedelli2014measuring}, gravity gradients \cite{snadden1998measurement, mcguirk2002sensitive, sorrentino2014sensitivity}, rotations \cite{gustavson1997precision, stockton2011absolute, dutta2016continuous}, gravitational redshift \cite{muller2010precision, roura2020gravitational}, and the fine-structure constant \cite{parker2018measurement}. The high sensitivity of the atomic interferometers has opened a promising direction to test the fundamental nature of gravity beyond the Newtonian limit, such as Einstein's equivalence principle \cite{dimopoulos2007testing, tarallo2014test, hartwig2015testing, zhou2015test, rosi2017quantum, geiger2018proposal}, the inverse law of gravitation \cite{yang2012test, biedermann2015testing}, gravitational wave detection \cite{dimopoulos2008atomic, graham2013new}, and dark energy \cite{hamilton2015atom, sabulsky2019experiment}.
Moreover, a promising scheme to test a nonclassical nature of gravity has recently been proposed \cite{bose2017spin, marletto2017gravitationally}.

The basic procedure of an atomic fountain is to prepare a superposition of the internal energy states of atoms and to split their paths depending on the internal states so that the wave packet in each trajectory acquires a different phase \cite{kasevich1991atomic, kasevich1992measurement}.
The phase difference induced by the external field can be detected by measuring the population of the internal states by applying Raman pulses in a Mach-Zehnder type of configuration ($\pi/2-\pi-\pi/2$). Since the size of the area enclosed by the two trajectories of the atoms determines the sensitivity of the interferometer, there have been many theoretical proposals and numerous experimental efforts to increase the area enclosed by the interferometer \cite{mcguirk2000large, muller2008atom, clade2009large, chiow2011102, asenbaum2017phase, kovachy2015quantum}.

Recently, a variant of the atomic fountain configuration, the so-called Stern-Gerlach (SG) matter-wave interferometer \cite{gerlach1922experimentelle, bohm2012quantum, wigner1963problem, machluf2013coherent, zimmermann2017t, margalit2018realization, amit2019t, margalit2019analysis}, has been proposed and experimentally implemented. 
Such type of interferometer employs magnetic fields to give a momentum kick to atoms instead of laser pulses inducing Raman transitions.
A SG interferometer requires an extreme accuracy of the field gradients to maintain coherence \cite{englert1988spin, schwinger1988spin, scully1989spin}. Remarkably, a recent state-of-the-art experiment \cite{amit2019t} has successfully implemented the interferometer.
Besides, the experiment has shown that the SG interferometer enables the atomic ensemble to have the phase difference with $T^3$ scaling, where $T$ is the total interferometer time. This provides a higher sensitivity to the acceleration induced by the external field than the conventional atomic fountain using Raman transitions, rendering $T^2$ scaling.
Along with this proposal, it has been shown that external fields instead of laser-pulses can be exploited to enhance the precision of matter-wave interferometers \cite{comparat2020limitations}.

Atomic interferometers can also be utilized for probing the gradient of an external field, i.e., when the acceleration varies by position. The precise measurement of field gradients is particularly important for testing fundamental physics such as the variation of the gravitation constant $G$ and the violation of the $1/r^2$ law \cite{damour1994string, yang2012test, biedermann2015testing}. Besides, it has various technical applications, including detecting subsurface mass anomalies \cite{butler1984microgravimetric, mikhailov2007tensor}, inertial navigation system \cite{lawrence2012modern}, underground structure detection \cite{romaides2001comparison}, and mineral exploration \cite{fischbach1986reanalysis}.

The effect of a field gradient appears in the conventional Mach-Zehnder type atomic interferometers because it causes a systematic misalignment of position and momentum, which eventually diminishes visibility \cite{prevedelli2014measuring, roura2014overcoming}. The obstacle can be overcome if a precise estimate of the gravity gradient is given \cite{roura2014overcoming,d2017canceling}.
An interesting way to employ an atomic interferometer for measuring the field gradient is to construct a butterfly configuration \cite{mcguirk2002sensitive, clauser1988ultra, marzlin1996state, canuel2006six, dubetsky2006atom, wu2007demonstration, tonyushkin2008selective, takase2008precision, stockton2011absolute, kleinert2015representation}.
The butterfly interferometer enables us to analyze the second-order effect precisely because the area diagram between two wave packets in space-time, which contribute to the first-order effect, cancel out \cite{zimmermann2019representation}.

In this paper, we introduce an SG interferometer in a butterfly configuration to measure a field gradient. By applying a magnetic field to split the atoms' trajectory depending on the magnetic moments of the internal atomic states, the interferometer can be closed up to zeroth-order in the field gradient strength. 
Applying the magnetic field provides a more flexible control of the interferometer paths than laser-pulses, resulting in a higher sensitivity in the phase difference. 
We compare the performance of the proposed interferometer with that of the conventional one. 
From this, we show that the phase difference can be efficiently increased by inducing a larger momentum splitting in the interferometer and using larger magnetic moments of atoms. 
We also verify that the interferometer maintains a high visibility even though the misalignment of the interferometer can occur in the higher orders of the field gradient strength.

\section{Conventional Butterfly interferometer}
Let us consider a quantum system exposed to an external field having a quadratic potential $V(\hat{z}) = -F_0 \hat{z} - \frac{1}{2}K \hat{z}^2$ in one dimension (1-D) with constants $F_0$ and $K$.
The Hamiltonian of the system is then described as
\begin{align}
    \hat{H}=\frac{\hat{p}^2}{2m}-F_0\hat{z}-\frac{1}{2}K\hat{z}^2.
\end{align}
Here, $\hat{z}$ and $\hat{p}$ are quantized position and momentum operators along the $z$ axis, satisfying the canonical commutation relation $[\hat{z}, \hat{p}] = i \hbar$.
When $K<0$, the last term is a harmonic potential that provides a restoration force to a stable point determined by $F_0$ and $K$. When $K>0$, a force from the last term pushes the atoms outward. 
For example, the Hamiltonian can describe the effect of gravity by setting $F_0=-mg$ and $K=m\Gamma$ with the gravitational acceleration $g$ and gravity gradient $\Gamma$.

\begin{figure}[t]
\centering
\includegraphics[width=0.45\textwidth]{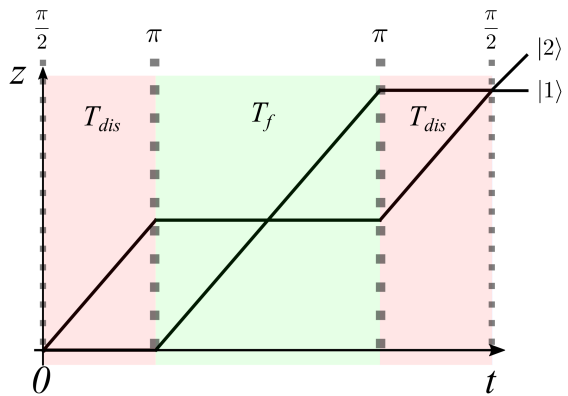}
\caption{Conventional butterfly interferometer in a free falling frame. The trajectories describe the mean position of each wave packet.}
\label{fig:conventional}
\end{figure}

We consider the problem of measuring the field gradient, given by the parameter $K$.
The conventional butterfly interferometer \cite{kleinert2015representation} measures  the constant $K$ as follows:
First, a $\pi/2$ pulse is applied to an atomic cloud prepared in the ground state $|1\rangle$ of the internal state to generate an equal superposition between the ground state $\ket{1}$ and the excited state $\ket{2}$.
By applying the pulse, the wave packet in the internal state $|2\rangle$ gains momentum, which leads to a momentum difference between the two trajectories $\delta p=\hbar k$ proportional to the wave number $k$ of the laser pulse.
Second, after waiting $T_\text{dis}$ for displacement of position, a $\pi$ pulse is applied to swap the momentum states of the two different trajectories.
Third, the wave packets freely evolve for a free evolution time $T_f$. 
Finally, we repeat the first two steps in the reverse way and measure the internal state on the basis $\{|1\rangle,|2\rangle\}$.
The procedure is illustrated in Fig.~\ref{fig:conventional}.
Since the trajectory of each wave packet is different, the phase accumulation is also different for each trajectory. 
One can find that the phase difference before the measurement is written as \cite{zimmermann2019representation}
\begin{align}\label{conv_phase}
    \delta \Phi_\text{c}=-\frac{KT^2 k}{32 m } \left(\frac{p_0}{m}T+\frac{\hbar k}{2m}T+\frac{F_0}{2m} T^2 \right),
\end{align}
where $T = 2 T_\text{dis} + T_f$ is the total interferometer time, 
and $p_0$ is the initial momentum of the wave packet.
Since the probabilities of obtaining outcomes $\{|1\rangle,|2\rangle\}$ are $P_1=(1+ C\cos \delta \Phi_c)/2$ and $P_2=(1- C\cos \delta \Phi_c)/2$, the parameter $K$ can be inferred by the population of each state.
Here, $C$ represents the visibility that quantifies the interference of the atomic interferometer.
We discuss the effect of the visibility in the following section.

One can easily verify that the interferometer is closed in the zeroth order of $K$ when the free evolution time satisfies
\begin{align}\label{close_condition}
    T_f=\frac{2m\delta z}{\delta p},
\end{align}
where $\delta z$ is the difference in position right before the free evolution.
In the case of the conventional butterfly configuration, the preparation time $T_\text{dis}$ for displacement has to obey  $T_\text{dis}=T_f/2$ to meet the condition Eq.~\eqref{close_condition} by noting that $\delta z=\delta p T_\text{dis}/m$.
Thus, the total time duration of the interferometer is given by $T =2 T_f$.
When we consider the gravitational acceleration and gravity gradient, we recover the phase difference from the gravity gradient, which is obtained in Refs. \cite{sorrentino2014sensitivity, kleinert2015representation}.

More specifically, Eq.~\eqref{conv_phase} presents that the phase difference is decomposed into two factors: (i) the velocity difference $\delta p/m$ of the two wave packets, which corresponds to the term in front of the parentheses and (ii) the difference between the initial and final positions of the wave packet, which corresponds to the terms inside of the parentheses.
Thus, the sensitivity of measuring $K$ can be improved by increasing the difference of the velocities of the wave packets or that of the initial and final position of the wave packet.
As recently pointed out in Ref. \cite{comparat2020limitations}, the laser-pulse-based atomic interferometer has a limitation of increasing the momentum difference of the wave packets while employing external fields to give a momentum difference may resolve this drawback.
As a remark, since the difference between the initial and final positions is required to be large, it is beneficial to have the direction of an initial velocity to be the same as that of the linear force.
For example, if the final position is the same as the initial one, which may arise in a fountain configuration, the phase difference of the two trajectories vanish.


\begin{figure}[!t]
\centering
\includegraphics[width=0.45\textwidth]{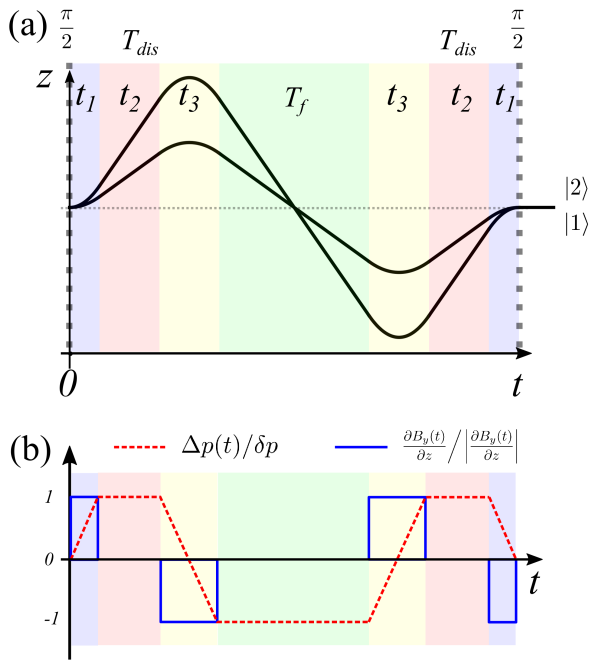}
\caption{SG butterfly interferometer in a free falling frame assuming $\hbar k \rightarrow 0$. (a) The trajectories describe the mean positions of the atomic states $\ket{1}$ and $\ket{2}$ having magnetic moments $\mu_1$ and $\mu_2$. (b) The momentum difference between the two wave packets, $\Delta p(t)=p_2(t)-p_1(t)$, and the magnetic gradient $\partial B_y(t)/\partial z$ that is applied to the atoms. Note that the two wave packets do not need to have the same direction of momentum.}
\label{fig:asym}
\end{figure}

\section{SG butterfly interferometer}
\subsection{Implementation of SG butterfly interferometer}
We propose a SG butterfly interferometer that exploits an external magnetic field, which is a variant of the conventional butterfly interferometer.
The configuration of the interferometer is illustrated in Fig. \ref{fig:asym}.
Such a configuration can be experimentally implemented by applying a magnetic field in the $y$-direction $\vec{B}(t) = (0, B_y(t), 0)$ having a gradient in the $z$-direction to give momentum kicks to wave packets differently based on their magnetic moments $\mu_1$ and $\mu_2$ (Fig.~\ref{fig:asym}) with $\mu_i=\mu_B g_{F_i} m_{F_i}$. Here, $\mu_B$, $g_{F_i}$, and $m_{F_i}$ denote the Bohr magneton, the Land\'{e} factor, and the magnetic quantum number along the $y$ axis, respectively.
Including the magnetic-field contribution, the force exerted on the atoms becomes time-dependent,
\begin{align}
    \hat{H}=\frac{\hat{p}^2}{2m}-F(t)\hat{z}-\frac{1}{2}K\hat{z}^2,
\end{align}
where $F(t)$ consists of a given linear force $F_0$ and the time-dependent magnetic force that we apply;
\begin{align}
    F(t) = F_0+\mu_1 \frac{\partial B_y(t)}{\partial z} \ket{1}\bra{1} + \mu_2 \frac{\partial B_y(t)}{\partial z} \ket{2}\bra{2},
\end{align}
where $\ket{1}$ and $\ket{2}$ are the internal quantum states having magnetic moments $\mu_1$ and $\mu_2$, respectively. Because of the difference between the magnetic moments, the wave packets experience different forces from the magnetic field. 

A detailed protocol for measuring the field gradient $K$ is given as follows: We prepare a quantum state in an equal superposition between two different internal states by applying a $\pi/2$ pulse
\begin{align}
|\Psi(0)\rangle=\frac{1}{\sqrt{2}}\left(|1\rangle|\psi_1(0)\rangle+|2\rangle|\psi_2(0)\rangle\right),
\end{align}
where the first and second kets represent the internal state and the spatial wave function of the atom, respectively. The wave packet in the internal state $|2\rangle$ obtains a momentum $\hbar k$ from the pulse.

We then apply magnetic field $\vec{B}(t)$ for $T_\text{dis}$. 
As the simplest configuration, we assume a sequence of the magnetic field,
\begin{align}\label{Bseq}
    B_y(t)=
    \begin{cases}
        b z & 0\leq t<t_1 \\ 
        0 & t_1\leq t < t_1+t_2 \\ 
        -b z & t_1+t_2\leq t<t_1+t_2+t_3,
    \end{cases}
\end{align}
where $b$ is the magnitude of the field gradient along the $z$ axis, and we denote $t_1, t_2, t_3$ as the time duration for each step composing the total time duration for displacement $T_\text{dis}=t_1+t_2+t_3$, as illustrated in Fig. \ref{fig:asym}.
Due to the magnetic field, the dynamics of the wave packets depend on their magnetic moments such that
\begin{align}
|\Psi(t)\rangle=\frac{1}{\sqrt{2}}\left(|1\rangle \hat{U}_1(t,0)|\psi_1(0)\rangle+|2\rangle\hat{U}_2(t,0)|\psi_2(0)\rangle\right),
\end{align}
where the unitary operators applied to the motion are written as \cite{zimmermann2019representation}
\begin{align}
    \hat{U}_i(t,0)=e^{i\Phi_i(t)}\hat{D}[z_i(t),p_i(t)]\hat{V}(t,0).
\end{align}
Here, $\Phi_i(t)$ is the phase accumulation on each trajectory, and the dynamics of the spatial wave function can be divided into the time-evolution operator $\hat{V}(t)=\exp\left[-\frac{it}{\hbar}\left(\frac{\hat{p}^2}{2m} - \frac{1}{2} K \hat{z}^2 \right)\right]$ without linear force and the displacement operator $\hat{D}[z_i(t),p_i(t)]=e^{i[p_i(t) \hat{z}- z_i(t) \hat{p}]/\hbar}$ due to the force exerted on the atoms. The displacement parameters $z_i$ and $p_i$ follow the dynamics of classical trajectories of the atoms. Explicitly, for a constant magnetic field in time, the dynamics is described by
\begin{widetext}
\begin{align}
z_i(t)&=\left[z_i(t_0)+\frac{F_i(t)}{m\omega^2}\right]  \cosh\omega(t-t_0)+\frac{p_i(t_0)}{m\omega}\sinh \omega(t-t_0)-\frac{F_i(t)}{m\omega^2}, \label{position}\\
p_i(t)&=\left[\frac{F_i(t)}{\omega}+m\omega z_i(t_0)\right] \sinh\omega(t-t_0)+p_i(t_0)\cosh\omega(t-t_0), \label{momentum}
\end{align}
\end{widetext}
where $t_0$ is the initial time, and we define $\omega\equiv \sqrt{K/m}$, and $F_i(t)=F_0+\mu_i \partial B_y(t)/\partial z$ is the force that each wave packet experiences.
The phase accumulation of each wave packet is written as
\begin{align}
    \Phi_i(t)=\frac{1}{2\hbar}\int_0^t d\tau z_i(\tau)F_i(\tau)+\phi_i,
\end{align}
where $\phi_i$ denotes the phase accumulation from $\pi/2$ pulses \cite{zimmermann2019representation}.
Since we focus on the time-dependent phase accumulation, we omit the phase $\phi_i$ from $\pi/2$ pulses throughout the paper.

After this magnetic-field sequence, assuming $\omega T \ll 1 $, the displacement of the position and momentum right before the free evolution, i.e., at $t=T_\text{dis}$, can be easily found by using Eqs.~\eqref{position} and \eqref{momentum}, and one can determine the free evolution time $T_f$ by the condition given by Eq.~\eqref{close_condition}.
After the free evolution time $T_f$, we apply the magnetic field in the reverse sequence to Eq.~\eqref{Bseq}, as illustrated in Fig.~\ref{fig:asym}. Finally, after another $\pi/2$ pulse, the internal states of the atoms are measured in the basis $\{|1\rangle,|2\rangle\}$.
As a result, the phase difference between two wave packets is written as
\begin{align}
\label{Eq:dPhi}
    \delta \Phi&=\Phi_2(T)-\Phi_1(T)+\frac{z_1(T)p_2(T)-z_2(T)p_1(T)}{2\hbar},
\end{align}
where the last term in Eq.~\eqref{Eq:dPhi} arises from the misalignment of the wave packets caused by nonzero constant $K$. 
As the misalignment not only leads to an additional phase but also causes a degradation of visibility of the interferometer, it is demanded to minimize it in practice.
We first focus on the phase difference $\delta \Phi$ and then discuss the effect of diminished visibility in Sec. \ref{sec:visibility}.
Note that in general, the phase difference depends on the initial position and momentum of the wave packets.


\begin{figure}[b]
\includegraphics[width=0.4\textwidth]{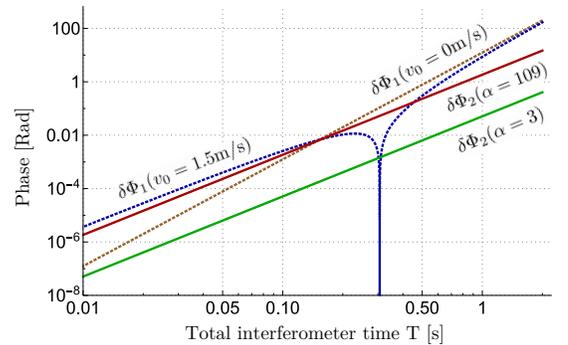}
\caption{Phase difference from (i) difference between the initial and final positions, $\delta\Phi_1=\big|\frac{K T^2}{32\hbar}\frac{\delta p}{m}\left(\frac{p_0}{m}T+\frac{F_0}{2m} T^2\right)\big|$, and (ii) a magnetic moment dependent term, $\delta\Phi_2=\big|\frac{K T^3}{32\hbar}\frac{\delta p^2}{m^2}\frac{(\mu_1+\mu_2)}{3(\mu_2-\mu_1)}\big|$ in Eq.~\eqref{Eq:Phase0}. Here, $v_0=p_0/m$ is the initial velocity. We assume that $\hbar k \rightarrow 0$. See the main text for details.}
\label{fig:plot1}
\end{figure}

\subsection{Magnetic-field gradient pulse regime}
Let us focus on an important limit that a magnetic-field gradient pulse is applied, i.e., we take a limit $t_1, t_3 \ll T_f$ with fixing the momentum transfer of the magnetic fields as
\begin{align}
    \label{Eq:Close1}
    (\mu_2-\mu_1)bt_1&=\delta p - \hbar k\\
    (\mu_2-\mu_1)bt_3&=2\delta p.\label{Eq:Close2}
\end{align}
Here, $\delta p$ represents the momentum difference between the two trajectories at $t=T_\text{dis}$, and $t_2$ is determined by the condition \eqref{close_condition} as $t_2=T_f/2$.
Under these conditions, the phase difference \eqref{Eq:dPhi} becomes
\begin{align}
    \label{Eq:Phase}
    \delta \Phi= &\left( \frac{\delta p}{\hbar k} \right) \delta \Phi_c + \frac{(\mu_1+\mu_2)}{(\mu_2-\mu_1)} \frac{K T^3 \delta p}{32 m^2 \hbar} \left( \frac{\delta p}{3} + \frac{\hbar k}{2} \right).
\end{align}
In this pulse regime, the role of the magnetic-field gradient pulse is similar to $\pi/2$ and $\pi$ pulses in the conventional butterfly configuration.
Especially by assuming $\delta p = \hbar k$ and $\mu_1 = -\mu_2$, the phase difference $\delta \Phi$ is exactly the same as that of the conventional butterfly interferometer, $\delta \Phi_c$. 
In this case, no magnetic field is applied during $0\leq t \leq t_1$ as we have $t_1 = 0$ from Eq.~\eqref{Eq:Close1}, and the magnetic field $t_2 \leq t \leq t_3$ has the same role as the $\pi$ pulse in the conventional butterfly setup.

When the momentum difference $\delta p$ is fixed, there is a trade-off relation between the difference of the magnetic moments $\mu_2-\mu_1$ and the impulse per magnetic moment from the magnetic field $b t_1$ and $b t_3$ given by Eqs. \eqref{Eq:Close1} and \eqref{Eq:Close2}.
Thus, in order to use the fact that the phase difference is proportional to $(\mu_2-\mu_1)^{-1}$, the strength of magnetic-field gradient is required to be large enough to satisfy the conditions Eqs. \eqref{Eq:Close1} and \eqref{Eq:Close2}.
Indeed, the underlying phase difference from the magnetic field is proportional to $(\mu_2^2-\mu_1^2)bdt$ $(dt\approx dt_1,dt_3)$, and it is simplified as Eq.~\eqref{Eq:Phase} due to the aforementioned trade-off relation.

We highlight that the magnetic-field gradient pulse can
provide an additional control in the momentum transfer $\delta p$ to increase the sensitivity of the interferometer compared with using the conventional laser pulses \cite{comparat2020limitations}.
One can observe that the first term of Eq.~\eqref{Eq:Phase} has the same form as in the conventional butterfly interferometer but scales linearly by increasing the momentum difference $\delta p$. Thus, when the magnetic field induces a larger momentum transfer $\delta p$ than that from the Raman transition $\hbar k$, the accumulated phase difference can be increased with the factor of $\delta p /(\hbar k)$.

Meanwhile, the second term in Eq.~\eqref{Eq:Phase}, which is magnetic moments dependent, is a distinct feature of the SG butterfly interferometer. In contrast with the first term or $\delta \Phi_c$, this term does not depend on the initial and final position of the atoms in the interferometer, thus can yield a nonvanishing phase difference even when the final position of the wave packet is the same as the initial position, which leads to $\delta \Phi_c =0$.
We also note that $\alpha \equiv (\mu_1 + \mu_2)/(\mu_2 - \mu_1)$ determines the prefactor of the magnetic-moment-dependent phase difference from the butterfly configuration of the SG interferometer.
To increase the factor $\alpha$, the magnitude of the magnetic moments $\mu_1$ and $\mu_2$ should be as large as possible, while keeping their difference $|\mu_2 - \mu_1|$ small. Note that $\alpha =0$ when $\mu_1 = -\mu_2$, in which case the interferometer has a symmetric geometry .
For example, a recent experiment on the atomic interferometer \cite{amit2019t} has implemented the $5^2 S_{1/2}$ manifold of $^{87} {\rm Rb}$ atom to generate superposition between $\ket{F=2, m_F = 2}$ and $\ket{F=2, m_F = 1}$. In this case, the prefactor is $\alpha = 3$.
This prefactor can be increased by employing an atom in superposition between higher magnetic moments. Rydberg atoms can be a promising option to realize this because they can have high quantum numbers of $n$, $l$, and $m$. Superposition between circular Rydberg states $\ket{55c} = \ket{n=55,l=54,m=54}$ and $\ket{56c} = \ket{n=56,l=55,m=55}$ recently realized in experiment \cite{palmer2019electric} can boost the factor to $\alpha = 109$, which is significantly higher than using the $5^2 S_{1/2}$ manifold of $^{87} {\rm Rb}$.

To focus on the effect of the magnetic field, let us consider the case $\hbar k \ll \delta p, p_0, F_0 T$ so that the initial and final $\pi/2$ pulses have a negligible role in the momentum transfer and the phase difference. We simply denote such a regime as $\hbar k \rightarrow 0$. A relevant physical situation would be when the atoms' internal states $\ket{1}$ and $\ket{2}$ undergo the Raman transition at radio frequencies \cite{margalit2019analysis}.
In this case, the phase is written as
\begin{align}\label{Eq:Phase0}
    \delta \Phi_{\hbar k\to0}=&-\frac{K T^2}{32\hbar}\frac{\delta p}{m}\bigg[\frac{p_0}{m}T+\frac{F_0}{2m} T^2 - \frac{\delta p}{m}\frac{(\mu_1+\mu_2)}{3(\mu_2-\mu_1)}T \bigg].
\end{align}

As an illustrative example, we have plotted in Fig.~\ref{fig:plot1} the phase differences in the SG butterfly interferometer induced by the gravitational field. As physical parameters, we take the experimentally feasible values of 
$m=1.42\times 10^{-25}\text{kg},~g=9.8\text{m}/\text{s}^2,~\Gamma=mK=3\times 10^{-6}\text{s}^{-2},~\delta v=\delta p/m=0.02 \text{m}/\text{s}$
based on Ref.~\cite{amit2019t}.
First of all, the figure shows that when the initial and final positions are the same, only the phase difference from the magnetic moment is nonvanishing.
Also, when Rydberg atoms are employed with $\alpha=109$, the phase difference from the magnetic moment can be large enough to increase the sensitivity.

\begin{figure}[t]
\centering
\includegraphics[width=0.4\textwidth]{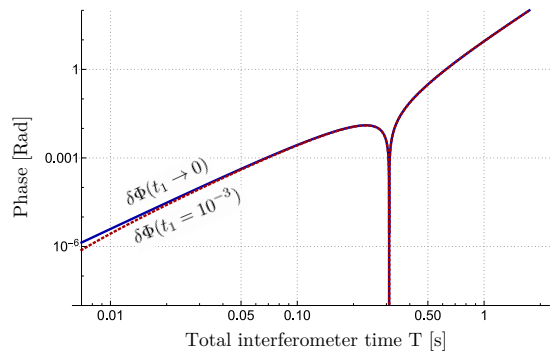}
\caption{Phase difference with a finite time duration for displacement fixing $\delta v=0.02 \text{m/s}$ \cite{amit2019t}. The magnitude of the gradient of magnetic field we need is $b=0.34\text{T/m}$ for $t_1=10^{-3}\text{s}$. The maximum separation is less than $0.01\text{m}$. Note that we have observed that the difference is very small even for larger $t_1$ such as $t_1=10^{-2}\text{s}$. We assume that $\hbar k \rightarrow 0$.}
\label{fig:plot2}
\end{figure}

In realistic experiments, the magnetic field in the displacement stage has to be large enough to attain the phase $\delta\Phi$ of the above limit.
Taking into account practical situations, we analyze the phase accumulation with a finite time duration of displacement instead of the pulse regime.
In this case, from the condition \eqref{close_condition}, the free evolution time is determined by $T_f=t_1+2t_2$.
We show the phase accumulation from finite time of displacement in Fig. \ref{fig:plot2}.
As shown in the figure, the finite time duration for displacement still leads to a phase difference similar to what we obtained in the limit $t_1\to 0$.
Since the time duration for displacement is not required to be extremely small, fixing the magnetic moment difference and the momentum difference in Eqs. \eqref{Eq:Close1} and \eqref{Eq:Close2}, one may employ a smaller magnetic-field gradient by increasing $t_1$ or $t_3$ in practice.


\subsection{Visibility}\label{sec:visibility}
Finally, let us discuss the degradation of the visibility due to the misalignment. The misalignment changes the probabilities of detecting $|1\rangle, |2\rangle$ as $P_{1}=(1+ C\cos \delta \Phi)/2$ and $P_{2}=(1- C\cos \delta \Phi)/2$,
where
\begin{align}
    C=|\bra{\psi_0(T)}\hat{V}^\dagger(T)\hat{D}(\Delta z(T), \Delta p(T))\hat{V}(T)\ket{\psi_0(T)}|
\end{align}
represents the visibility.
Here, $\Delta z(T)=z_2(T)-z_1(T)$ and $\Delta p(T)=p_2(T)-p_1(T)$.
Unless $\Delta z(T)=\Delta p(T)=0$, the visibility of the signal is strictly less than 1.
In the butterfly interferometer case with a pulse regime, up to order of $K$, we have
\begin{align}
    \Delta z(T)\approx \frac{K T^3 \delta p}{32m^2}, ~~~~ \Delta p(T)\approx 0.
\end{align}
In the case of Mach-Zehnder type of atomic interferometer, the misalignment is given by
\begin{align}
    \Delta z(T)\approx \frac{KT^3\delta p}{8m^2},~~~~   \Delta p(T)\approx \frac{KT^2\delta p}{4m}.
\end{align}

\begin{figure}[t]
\includegraphics[width=0.4\textwidth]{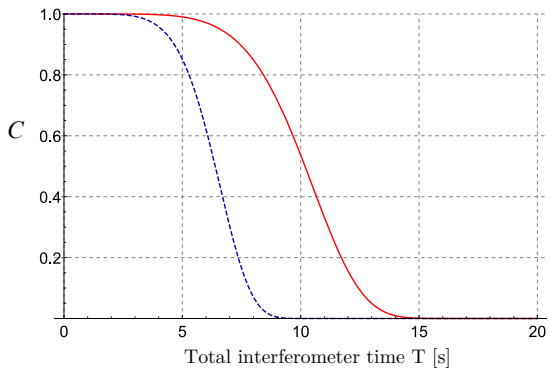}
\caption{Visibility of atomic interferometers using  $^{87}\text{Rb}$ atoms with the position uncertainty $\sigma_z=200 \mu\text{m}$ and the velocity uncertainty $\sigma_{p_z}/m=0.44 \text{mm}/\text{s}$, which is achieved with a temperature $T_\text{eff}=2\text{nK}$ \cite{roura2014overcoming}. The solid curve represents the visibility of the butterfly configuration and the dashed curve represents that of the Mach-Zehnder configuration.}
\label{fig:vis}
\end{figure}

To analyze the effect of the misalignment, let us assume the initial wave packet to be Gaussian, with the uncertainty of the position and momentum along the $z$ axis denoted $\sigma_z$ and $\sigma_{p_z}$. We assume that the correlation of the position and momentum is zero for simplicity.
In this case, the visibility $C$ for a pure state can be simplified as 
\begin{align}
C=|\langle\psi(0)|\hat{D}(\Delta z', \Delta p_z')|\psi(0)\rangle|=\exp\left[-\frac{\Delta z'^2}{8\sigma_{z}^2}-\frac{\Delta p_z'^2}{8\sigma_{p_z}^2}\right],
\end{align}
and for a mixed initial state $\hat{\rho}(0)$,
\begin{align}
C=|\text{Tr}[\hat{D}(\Delta z', \Delta p_z')\hat{\rho}_0]|=\exp\left[-\frac{\sigma_z^2 \Delta p_z'^2+\sigma_{p_z}^2 \Delta z'^2}{2\hbar^2}\right],
\end{align}
where $\Delta z'=\Delta z \cosh \omega T-\frac{\Delta p_z}{m\omega}\sinh \omega T$ and 
$\Delta p_z'=\Delta p_z\cosh \omega T+m\omega\Delta z \sinh\omega T$.
Figure \ref{fig:vis} shows the visibility of an example with different interferometer time $T$.
It shows that the visibility is approximately $1$ for $5~\text{seconds}$ and $3~\text{seconds}$ for the butterfly configuration and the Mach-Zehnder type interferometer, respectively.
Thus, the butterfly configuration is more robust to the misalignment.

When the visibility is degraded by the misalignment, one may overcome this obstacle by adjusting experimental parameters more elaborately \cite{roura2014overcoming, roura2017circumventing}.
For example, for an atomic interferometer based on laser pulses, by changing the frequency of the laser depending on the gravity gradient one can align atoms' trajectories to reduce the amount of the misalignment.
The same technique can be applied by changing the magnetic-field gradient accordingly.

\hfill

\section{Conclusion}
We have introduced a variant of a butterfly interferometer that enables a precise measurement of the field gradient. In our SG butterfly configuration, the atoms are prepared in superposition between the internal states having different magnetic moments, so that the external magnetic field can split the trajectories in the time-dependent manner.

The proposed interferometer is flexible in controlling the momentum transfer of the atoms by using the external magnetic field, while it implements the conventional butterfly configuration as a special case. We have shown that the phase difference between the different atomic trajectories arising from the field gradient can be decomposed into magnetic-moment-independent and -dependent parts. The magnetic moment independent phase can be compared with the phase difference in the conventional butterfly interferometer, but can linearly increase by the momentum difference of the interferometer $\delta p$ induced by the external magnetic field. The magnetic-moment-dependent phase provides a distinct feature of the SG butterfly interferometer scaling in $T^3$ and can be increased by enlarging the magnetic quantum number of the internal atomic state.

We have also investigated 
the degradation of the visibility when the interferometer is not perfectly closed due to the field gradient. We have demonstrated that the robustness of the interferometer against the misalignment is comparable to or better than the Mach-Zehnder type of atomic interferometer.


\begin{acknowledgements}
C.O. and L.J. acknowledge support from the ARL-CDQI (W911NF-15-2-0067), ARO (W911NF-18-1-0020, W911NF-18-1-0212), ARO MURI (W911NF-16-1-0349), AFOSR MURI (FA9550-15-1-0015, FA9550-19-1-0399), DOE (DE-SC0019406), NSF (EFMA-1640959, OMA-1936118), and the Packard Foundation (2013-39273).
The work of H.K. and M.S.K. was supported by the KIST Open Research Program.
M.S.K. acknowledges the visiting professorship at KIAS where this work was initially discussed.
The work of M.S.K. was supported by the QuantERA ERA-NET Cofund in Quantum Technologies implemented within the European Union's Horizon 2020 Programme and EPSRC (EP/R044082/1).
\end{acknowledgements}

\bibliography{reference}

\end{document}